\documentclass[10pt,conference]{IEEEtran}
\IEEEoverridecommandlockouts

\usepackage[final]{changes}  
\usepackage{cite}
\usepackage{amsmath,amssymb,amsfonts}
\usepackage{algorithmic}
\usepackage{graphicx}
\usepackage{textcomp}
\usepackage{xcolor}
\usepackage{balance} 
\usepackage{url} 
\usepackage{textcomp}
\usepackage{booktabs}
\usepackage{hyperref}
\usepackage{enumitem}
\usepackage{multirow}
\usepackage{subfig}
\usepackage{listings}
\usepackage{color}
\usepackage{threeparttable}
\usepackage{array}
\usepackage{diagbox}
\usepackage{graphicx}
\usepackage{verbatim}
\usepackage{url}
\usepackage{soul}
\soulregister\cite7
\soulregister\ref7
\usepackage[multiple]{footmisc}
\usepackage{graphicx}
\usepackage{multirow}
\usepackage{hhline}
\usepackage{float}
\usepackage{array,booktabs}
\usepackage{marvosym} 
\usepackage{pifont}
\usepackage{tcolorbox}
\usepackage{bm}
\usepackage[ruled,linesnumbered]{algorithm2e}
\usepackage{hyperref}

\definecolor{diffstart}{RGB}{173, 175, 177}
\definecolor{diffincl}{RGB}{0, 110, 0}
\definecolor{diffrem}{RGB}{255, 50, 0}
\definecolor{orange}{RGB}{255, 165, 0}
\definecolor{javapurple}{rgb}{0.5,0,0.35} 

\newcommand{\find}[1]{
\begin{tcolorbox}[leftrule=0.5mm,rightrule=0.5mm, toprule=0.5mm,bottomrule=0.5mm,left=2pt,right=2pt,top=2pt,bottom=2pt]
#1
\end{tcolorbox}
}

\newcommand{\appname}{{\sc TestUpdater}\xspace}
\newcommand{\appnamebold}{{\sc \textbf{TestUpdater}}\xspace}

\newcommand{\datasetname}{{\sc Updates4J}\xspace}
\newcommand{\datasetnamebold}{{\sc \textbf{Updates4J}}\xspace}

\begin{document}

\title{Unit Test Update through LLM-Driven Context Collection and Error-Type-Aware Refinement}
\author{%
  \IEEEauthorblockN{%
    Yuanhe Zhang,
    Zhiquan Yang,
    Shengyi Pan,
    Zhongxin Liu\IEEEauthorrefmark{1}%
  }%
  \IEEEauthorblockA{The State Key Laboratory of Blockchain and Data Security, Zhejiang University, Hangzhou, China\\
  \{yuanhezhang, zhiquanyang, shengyi.pan, liu\_zx\}@zju.edu.cn
  \thanks{\IEEEauthorrefmark{1} Corresponding author.}}
}

\maketitle

\begin{abstract}
Unit testing is critical for ensuring software quality and software system stability. 
The current practice of manually maintaining unit tests suffers from low efficiency and the risk of delayed or overlooked fixes. 
Therefore, an automated approach is required to instantly update unit tests, with the capability to both repair and enhance unit tests.
However, existing automated test maintenance methods primarily focus on repairing broken tests, neglecting the scenario of enhancing existing tests to verify new functionality.
Meanwhile, due to their reliance on rule-based context collection and the lack of verification mechanisms, existing approaches struggle to handle complex code changes and often produce test cases with low correctness.

To address these challenges, we propose \appnamebold, a novel Large Language Model (LLM) based approach that enables automated just-in-time test updates in response to production code changes. 
By emulating the reasoning process of developers, \appnamebold first leverages the LLM to analyze code changes and identify relevant context, which it then extracts and filters.
This LLM-driven context collector can flexibly gather accurate and sufficient context, enabling better handling of complex code changes.
Then, through carefully designed prompts, \appnamebold guides the LLM step by step to handle various types of code changes and introduce new dependencies, enabling both the repair of broken tests and the enhancement of tests.
Finally, emulating the debugging process, we introduce an error-type-aware iterative refinement mechanism that executes the LLM-updated tests and repairs failures, which significantly improves the overall correctness of test updates.

Since existing test repair datasets lack scenarios of test enhancement, we further construct a new benchmark, \datasetnamebold, with 195 real-world samples from 7 projects, enabling execution-based evaluation of test updates.
Experimental results show that \appnamebold achieves a compilation pass rate of 94.4\% and a test pass rate of 86.7\%, outperforming the state-of-the-art method \textsc{Synter} by 15.9\% and 20.0\%, respectively.
Furthermore, \appnamebold exhibits 12.9\% higher branch coverage and 15.2\% greater line coverage than \textsc{Synter}. 

\end{abstract}

\section{Introduction}
\label{intro}
Unit testing is a fundamental practice in modern software development~\cite{olan2003unit}. 
It verifies the functionality of individual components within a codebase, helping developers detect regressions early in the development cycle and maintain system stability~\cite{runeson2006survey,skoglund2004case}.
However, as software evolves, unit tests that are not updated accordingly can become outdated and incompatible with changes in production code, potentially resulting in compilation or runtime failures. 
Moreover, outdated unit tests fail to cover the newly introduced functionalities in the updated codebase, which can lead to undetected bugs and reduced test coverage~\cite{lubsen2009coevolution}.
Therefore, keeping unit tests up to date is essential for maintaining the quality and effectiveness of test suites~\cite{shamshiri2015automated,hurdugaci2012aiding}.

Manually updating unit tests to reflect changes in production code, however, is a time-consuming and error-prone process~\cite{kasurinen2010software,widder2019conceptual}. 
This challenge is pronounced considering large and complex projects, where frequent updates require continuous maintenance of tests. 
Therefore, automatically updating unit tests is crucial to reduce manual efforts and improve the efficiency and accuracy of test maintenance.

Early research efforts propose rule-based approaches to automatically repair outdated tests~\cite{daniel2011reassert,daniel2010test,xu2014using}.
However, these approaches depend on predefined rules or patterns, which fail to cover all possible test repair scenarios. 
With the advancement of deep learning, some researchers fine-tune pre-trained models (PTMs) to perform test repair and update, such as \textsc{Ceprot}~\cite{ceprot} and \textsc{TaRGET}~\cite{target}.
Recently, researchers have explored using large language models (LLMs) to automatically repair broken unit tests.
Liu \emph{et al.}~\cite{synter} introduce the first LLM-based approach (namely \textsc{Synter}) that uses GPT-4~\cite{openai2024gpt4technicalreport} to repair obsolete test cases.
\textsc{Synter} enhances LLM's ability by collecting and reranking test-repair-oriented contexts.

However, the existing approaches face the following three challenges:

\textbf{\textit{Challenge 1:} Test enhancement for new functionality.}
The real-world software evolution requires both test repair and test enhancement.
Automatically enhancing test cases to verify new functionality is challenging because it requires correctly determining how the new functionality should be verified and collecting the relevant context necessary to implement the corresponding test logic.
Such capabilities go beyond simple syntactic or signature changes and demand a deep semantic understanding of the codebase.
Tools like \textsc{TaRGET} and \textsc{Synter}, which focus on repairing broken tests due to syntactic or signature changes~\cite{target, synter}, fall short in scenarios requiring test enhancement to reflect new behavioral logic.

\textbf{\textit{Challenge 2:} Context collecting for complex code changes.}
A key challenge in test updates is to accurately capture sufficient context in complex code changes.
The context may include method dependencies, class definitions, and related variables, all of which help the LLM understand code changes and produce more effective results.
Existing methods rely on rule-based context extraction strategies, which fail to generalize beyond simple code edits and struggle to handle complex code changes.
In our evaluation, this challenge was evident in \textsc{Synter}, where at least 16.9\% of the updated test cases failed to compile due to missing or inadequate context.

\textbf{\textit{Challenge 3:} Correctness assurance after the LLM updates the test.}
Another major challenge in test update lies in how to correct erroneous outputs after the test has been updated by the LLM.
Although providing sufficient context can often enable the LLM to perform reasonable updates, it may still make unnecessary modifications or introduce undeclared dependencies.
As observed by Yuan \emph{et al.}~\cite{yuan2024evaluating}, although most LLM-generated tests are syntactically valid, a large portion still suffer from compilation or execution errors.
Existing approaches typically lack mechanisms for validating and refining the LLM’s output, which requires manual debugging and limiting the effectiveness.

To address \textit{\textbf{Challenge 1}}, we propose \appname, an LLM-based approach that can not only repair but also enhance tests (e.g., verifying new functionality) instantly in response to production code changes, a process we refer to as \textit{just-in-time test update}. 
We design carefully crafted prompts to guide the LLM step by step, enabling it to reason about behavioral changes and introduce new dependencies when necessary.
For \textit{\textbf{Challenge 2}}, in contrast to prior rule-based context collectors, we ask the LLM to mimic the reasoning process of developers to determine which classes or methods should be referenced when updating tests, and retrieve them through a language server~\cite{lsp}.
This design enables flexible and accurate dependency resolution, allowing \appname to handle a wide variety of code change scenarios.
Additionally, we introduce another important but previously overlooked context source: the predefined variables in the test class, which LLMs can use directly.
For \textit{\textbf{Challenge 3}}, we innovatively introduce an error-type-aware iterative refinement mechanism as a post-processing module to further enhance the tests updated by the LLM. 
Inspired by developers’ debugging practices, after executing the LLM-updated test, \appname detects and classifies errors (e.g., missing symbol, type mismatch, assertion failure) and maps each type to a pre-defined refinement strategy.
Unlike prior single-shot generation approaches, our iterative repair strategy effectively guarantees the reliability of LLM-updated tests.

Furthermore, existing test repair datasets often lack scenarios involving test enhancement.
To evaluate the effectiveness of our approach, we construct a new dataset \datasetname by collecting samples of method-level production-test co-evolution from the open-source Java projects.
\datasetname contains 195 samples from 7 different projects, covering real-world scenarios of software evolution and supporting execution-based evaluation.
Based on \datasetname, we assess the performance of \appname in terms of \textit{compilation pass rate}, \textit{test pass rate}, and \textit{test coverage} with three different LLMs: \texttt{GPT-4.1}, \texttt{Llama-3.3-70B-Instruct}, and \texttt{DeepSeek-V3}.
The best result achieved by \appname yields a 94.4\% compilation pass rate and an 86.7\% test pass rate, representing improvements of 15.9\% and 20.0\% over the state-of-the-art (SOTA) method \textsc{Synter}\cite{synter}, respectively.
In terms of test coverage, \appname achieves an overall branch coverage of 46.0\% and a line coverage of 69.1\%, surpassing \textsc{Synter} by 12.9\% and 15.2\%, respectively.
Furthermore, we validate the effectiveness of key components in \appname through an ablation study.

In summary, we make the following contributions in this paper:
\begin{itemize}[leftmargin=*]
    \item \textbf{\appnamebold: A novel LLM-based test update approach that goes beyond test repair.} 
    \appname can not only repair but also enhance the outdated unit tests.
    By dynamically combining the reasoning ability of LLMs with language servers and compiler feedback, \appname facilitates the algorithmic emulation of developers’ behavior in test update, and further formalizes this process into three essential steps: context collection, test generation, and iterative refinement.
    We open-source our replication package for follow-up works~\footnote{https://github.com/ZJU-CTAG/TestUpdater}.
    
    \item \textbf{\datasetnamebold: A test update evaluation dataset.}
    We construct a new dataset \datasetname to fill the gap in existing benchmarks that focus solely on test repair, lacking coverage of test enhancement scenarios. 
    \datasetname consists of 195 samples from 7 open-source Java projects, capturing real-world test update scenarios and enabling execution-based evaluations of automated test update approaches.
        
    \item \textbf{An empirical evaluation of \appnamebold on \datasetnamebold.}
    We conduct extensive experiments to evaluate the effectiveness of our approach. 
    The experimental results demonstrate that \appname can effectively update outdated unit test cases and outperforms the LLM-based test repair method \textsc{Synter}~\cite{synter}, and the fine-tune methods \textsc{Ceprot}~\cite{ceprot} and \textsc{TaRGET}~\cite{target}.
\end{itemize}
\section{Motivation}

\begin{figure*}[htbp]
  \centering
  \includegraphics[width=1.0\textwidth]{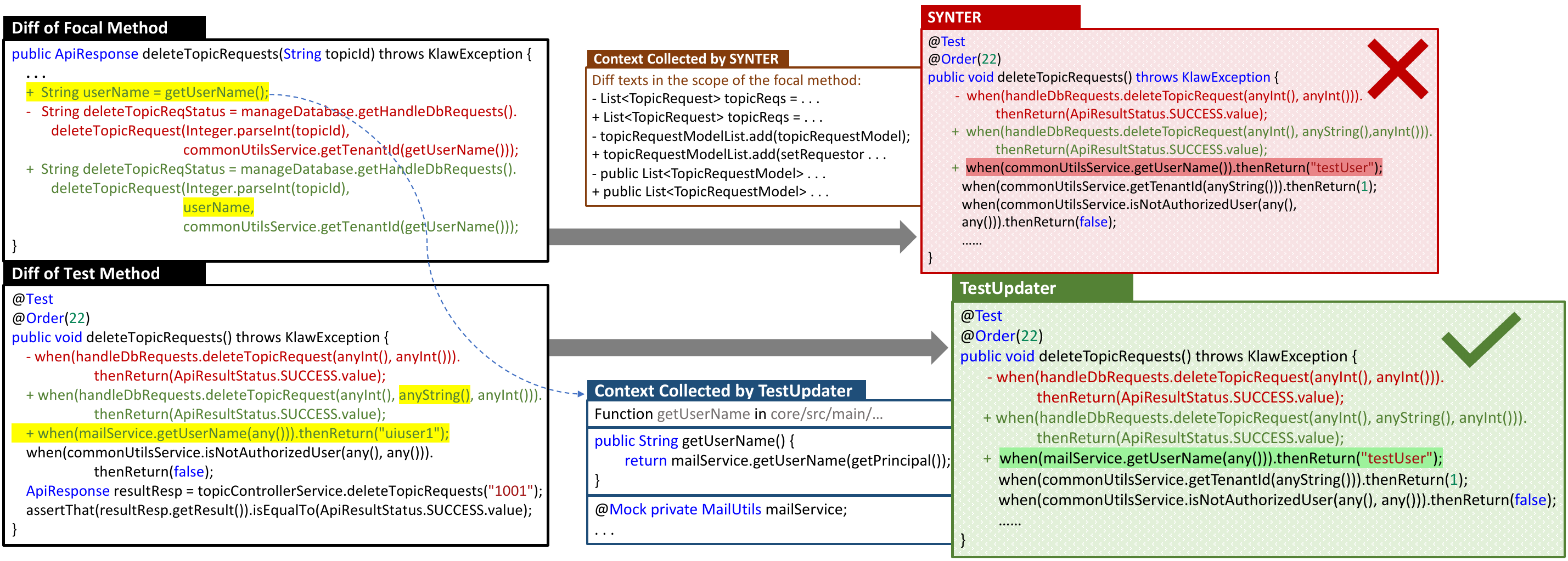}
  \caption{A Motivation Example For Context Collection from commit \textit{0f5599} in Project \textit{Aiven-Open/klaw}}
  \label{fig:motivation}
  \vspace{-10pt}
\end{figure*}

\begin{figure*}[htbp]
  \centering
  \includegraphics[width=0.98\textwidth]{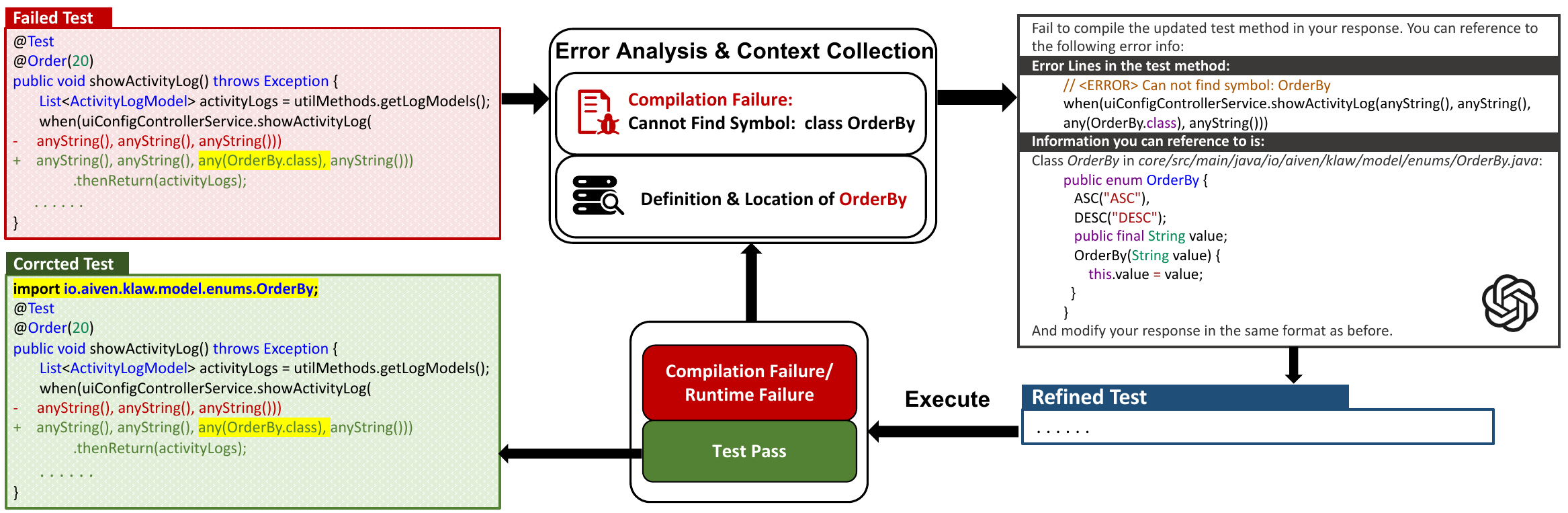}
  \caption{A Motivation Example for Error-Type-Aware Iterative Refinement from commit \textit{32e27ee} in Project \textit{Aiven-Open/klaw}}
  \label{fig:motivation_iter}
\end{figure*}

Fig.~\ref{fig:motivation} presents a production-test co-update example.
The method \texttt{deleteTopicRequest()} invoked in the focal method \texttt{deleteTopicRequests()} is updated to include a second parameter, \texttt{userName}. 
Consequently, the focal method is modified to invoke \texttt{getUserName()}, which uses \texttt{mailService.getUserName()} for the variable \texttt{userName}.
To accommodate the update of production code, corresponding adjustments are made in the test method to 
1)~update the mock of \texttt{deleteTopicRequest()} to include the additional \texttt{userName} parameter.
2)~mock the additional parameter with \texttt{mailService.getUserName()}. 

The test updated by \textsc{Synter} incorrectly mocks the username with \texttt{commonUtilsService.getUserName()}, which results in compilation failures (highlighted in red). 
The reason is that through rule-based algorithms, \textsc{Synter} only collects three diff texts in the focal class, fails to cover the necessary context of \texttt{mailService.getUserName()} within the \texttt{getUserName()} method.
Therefore, without knowing mock \texttt{mailService.getUserName()} for username, the LLM uses \texttt{commonUtilsService.getUserName()} just as the other mock behaviors in the test.

To accurately locate the required context, we observe that developers often start by examining the definitions of relevant functions and classes when updating tests (\texttt{getUsername()} in this case). 
However, rule-based algorithms are unable to replicate the reasoning process used by experienced developers because they lack the flexibility to understand complex relationships and contextual nuances that go beyond predefined patterns.
Given that LLMs possess human-like capabilities in code understanding and reasoning, we propose leveraging LLMs to identify the required context.
To this end, we employ an LLM-driven context collector to accurately locate the critical context—the definition of the \texttt{getUsername()} function. 
Combined with the mock instance \texttt{mailService} defined in the test class, \appname collects sufficient context and successfully performs the test update.

Another motivation is shown in Fig.~\ref{fig:motivation_iter}.
The red box in the figure highlights a failed test case where the LLM used \texttt{any(OrderBy.Class)} to mock a new parameter, but the compilation failed due to missing relevant dependencies.
This issue arises because the LLM did not obtain the location for \texttt{OrderBy} or failed to correctly import the dependency based on location information.
A simple idea to address this is to invoke the LLM again with the specific information about the \texttt{OrderBy} class.
The challenge lies in determining which information needs to be extracted based on the type of error.
To this end, we introduce an error-type aware iterative repair mechanism that extracts context and assists the LLM in fixing the test, as shown in the figure.
\section{Approach}
\label{sec:approach}
In this section, we first introduce the overall framework of \appname. 
Then, we describe each key step in detail, including \emph{context collection}, \emph{updated test generation}, and \emph{iterative refinement}.

\subsection{Overview}
\begin{figure*}[htbp]
  \centering
  \includegraphics[width=1.0\textwidth]{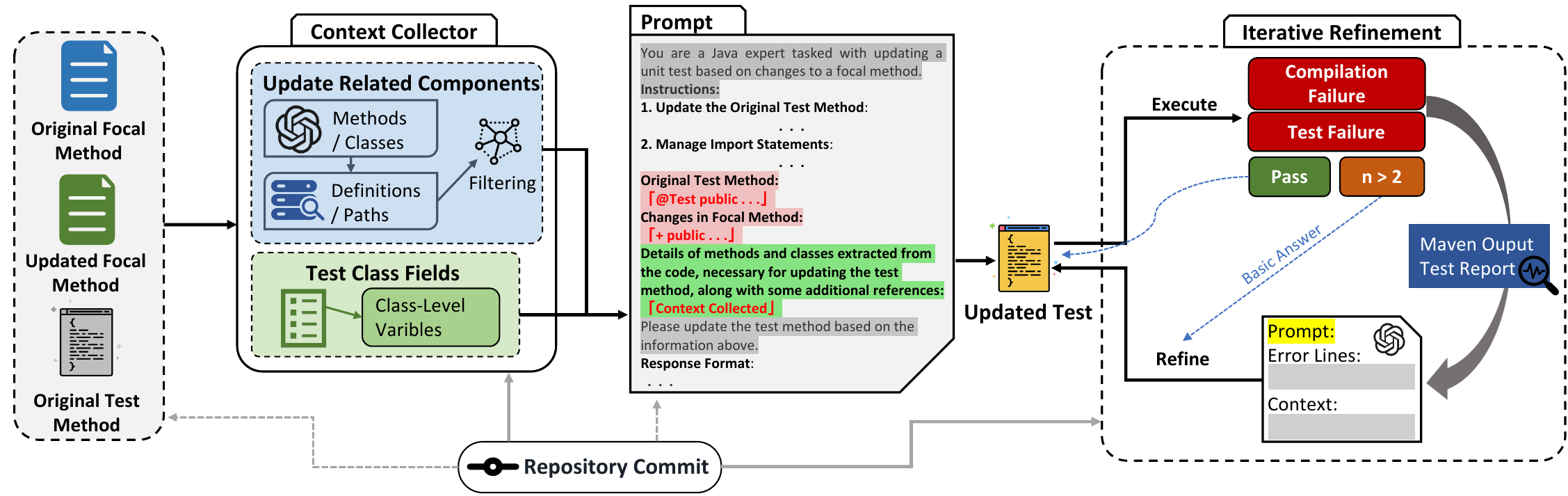}
  \caption{Overall Framework of \appname}
  \label{fig:framework}
\end{figure*}

To make it clear, in the rest of this paper, we use $T_o$ for the original test method, $T_u$ for the updated test method, $F_o$ and $F_u$ for the original and updated focal method, respectively.
We define our approach as follows.
\[
\text{\appname}(F_o, F_u, T_o, Ctx, Err) \rightarrow T_u
\]
where $Ctx$ represents the collected context and $Err$ represents the error feedback from execution.
The goal of \appname is to instantly update the associated test method according to the changes between $F_o$ and $F_u$, and generate $T_u$ that successfully compiles and passes against $F_u$.
The overall framework of \appname is illustrated in Fig.~\ref{fig:framework}. 
The entire process can be generally divided into three stages:

\begin{enumerate}[leftmargin=*]
    \item \textbf{Context Collection}:
    \appname initiates the update process through the LLM-driven context collector, which gathers the following contexts:
    1. Update Related Components: methods and classes that may be necessary for the test update, including their definitions and locations.
    2. Test Class Fields: all variables declared at the class level within the test class.
    
    \item \textbf{Updated Test Generation}:
    Based on the collected context, \appname constructs a structured prompt and invokes the LLM to automatically generate the updated test cases and introduce the required dependencies.
    
    \item \textbf{Iterative Refinement}:
    \appname compiles and runs the LLM-updated test cases.
    If compilation or test failures occur, \appname performs targeted context gathering according to the error types and invokes the LLM to perform refinements.
    The iterative validate-repair loop continues until the test cases compile and pass, or a preset iteration limit is reached, with fallback to minimal modifications to ensure correctness.
\end{enumerate}

\subsection{Context Collection}
\label{sec:context}
Context is crucial for enabling LLMs to generate correct and relevant test updates while reducing hallucinations.
Unlike prior methods that statically extract diff hunks or local signatures, we leverage the LLM to identify dependent methods and classes by reasoning over code changes, and then retrieve their definitions and locations by using a language server. 
Additionally, we introduce an additional type of context that has been overlooked by prior work - the predefined variables in the test class.
Overall, we categorize these contexts into two types: \textit{update related components} and \textit{test class fields}.

\textbf{\textit{Update Related Components.}}
In the actual process of test update, developers usually need to consult the definitions and locations of related methods and classes. 
By analyzing their definitions, developers can better understand their functionality, thereby reasonably mocking new objects, adjusting parameter configurations, or adding test logic. 
By identifying the locations, developers can determine the necessary dependencies to import, which ensures the correct compilation and execution of the updated test cases.

\begin{figure}[h]
  \centering
  \includegraphics[width=0.45\textwidth]{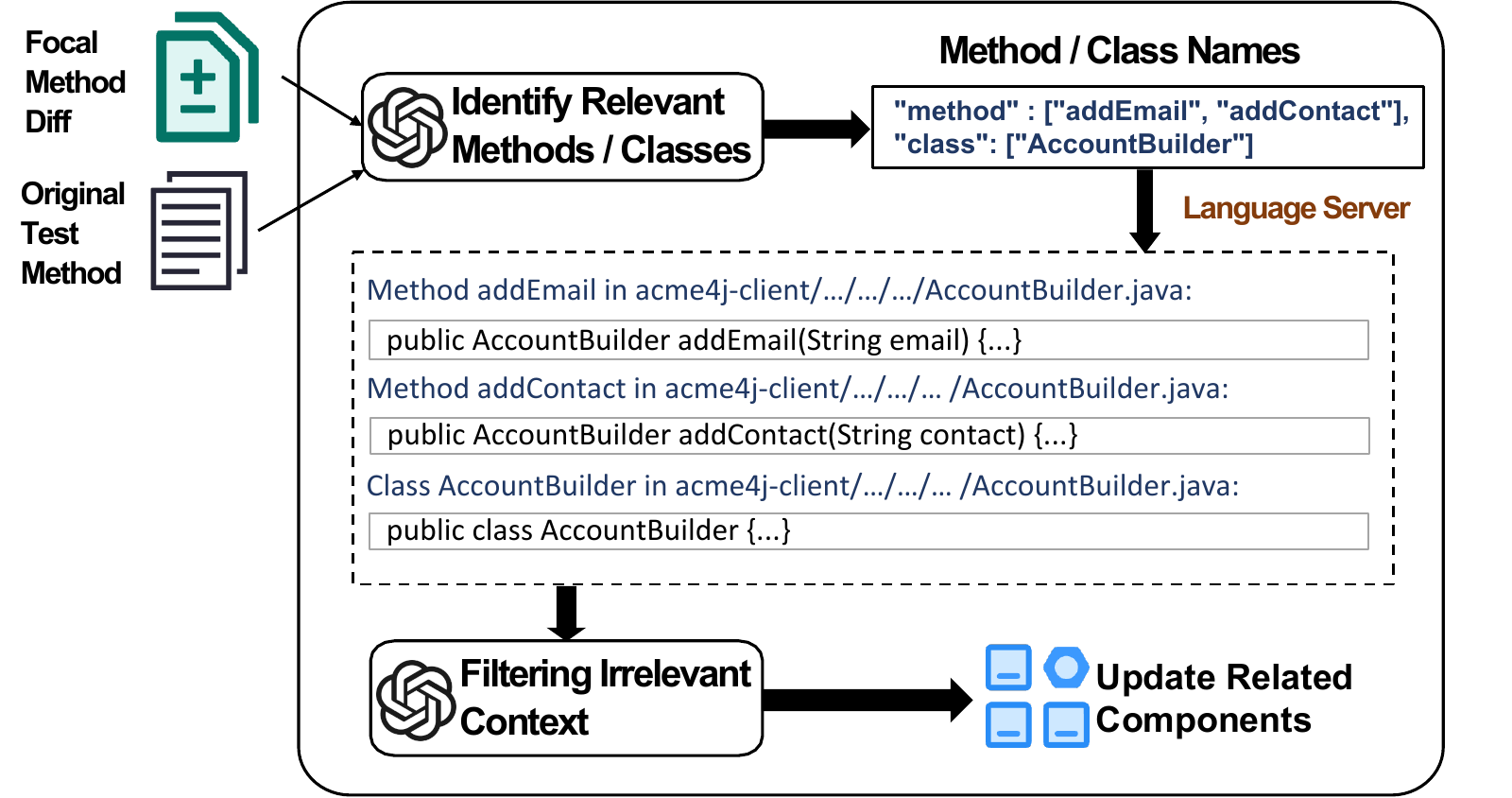}
  \caption{The process for collecting Update Related Components.}
  \label{fig:URC}
\end{figure}

Therefore, we design the Update Related Components context, which contains definitions and locations of relevant methods or classes. 
These methods and classes are typically referenced directly or indirectly in the updated test cases.
The whole process is illustrated in Fig.~\ref{fig:URC}, where an LLM mimics the developer’s reasoning to determine the necessary information and further utilizes tools (e.g., a language server, static analysis tools) to extract the required data.

The first step is to identify the names of the methods or classes that may be involved in the update. 
This process relies on a deep understanding of code semantics~\cite{sun2024method}, a level of comprehension that rule-based methods employed in existing works often fail to achieve.
Considering the strong reasoning capability of LLMs and their astonishing performance in multiple code understanding tasks~\cite{xu2022systematic} and automated software engineering tasks~\cite{hou2024large}, \appname leverages the LLM to analyze changes in the focal method and automatically identify the needed methods and classes. 
Specifically, we construct the prompt (c.f. our replication package for the detailed prompt) using chain-of-thought techniques~\cite{wei2022chain} with the unified diff of $F_o$ and $F_u$, and $T_o$ as inputs.
To ease the subsequent processing, we instruct the LLM to return the identification results in JSON format.

After identifying the methods and class names, the next step is to obtain their definitions and locations.
\appname leverages a Language Server~\cite{lsp} to locate the definitions of relevant methods and classes, and parse the code into Abstract Syntax Trees (ASTs)~\cite{ast} to extract the corresponding code snippets. 
Language Servers are often integrated into Integrated Development Environments (IDEs) to assist programmers in efficient development, which analyzes the entire codebase to provide advanced language features such as code completion, syntax checking, go-to-definition, and reference finding.

The raw definitions of methods and classes extracted from the codebase cannot be directly provided to the LLM, as they often include extraneous information, such as unrelated methods or variables. 
This irrelevant content introduces additional overhead and may mislead the LLM during the test update process. 
To address this, \appname directs the LLM to filter out unnecessary context from the raw definitions.
Meanwhile, \appname explicitly collects and preserves the source file paths of these methods and classes, enabling the LLM to accurately generate the required import statements.

\textbf{\textit{Test Class Fields.}}
In practical software development, developers frequently reuse class-level variables defined in test classes when updating unit test methods~\cite{zhu2024exploring}, which promotes consistency and reduces code redundancy. 
These variables usually contain instances of the classes under test, mocked dependencies, and configuration constants.
Motivated by this observation, \appname also collects Test Class Fields—all class-level variables defined within the test class, as a supplementary. 
Test Class Fields can prevent the LLM from unnecessarily mocking already defined variables, significantly reducing errors introduced by additional test logic.

\subsection{Updated Test Generation}
After collecting the contexts, the next step is to invoke the LLM to update the test cases. 
This phase consists of two key components: \textit{prompt construction} and \textit{post-processing}.

\textbf{\textit{Prompt Construction.}}
To enable the LLM to effectively perform both test repair and enhancement, we design a structured prompt template using the Markdown format. 
The prompt includes the following three inputs: 
(1) the unified diff of the focal method, 
(2) the original test method,
(3) the collected context. 
As shown in Fig.~\ref{fig:framework}, the prompt first sets the LLM’s role as a Java expert. 
Then, through step-by-step instructions, it guides the model to accomplish two tasks: \textit{update the original test method} and \textit{introduce required dependencies}.

Regarding the \textit{update original test method} part, the prompt provides five clear and actionable instructions to help the LLM repair and enhance the test method based on the changes in the focal method:
\begin{enumerate}[leftmargin=*]
    \item Update the test method using the provided context to align with changes in the focal method.
    \item Ensure the updated test correctly validates the new \deleted{behavior} logic.
    \item If the focal method introduces new functionality, generate new test logic accordingly.
    \item For any new parameters, mock the required objects or use default values.
    \item If the core functionality of the focal method remains unchanged, only repair the original test.
\end{enumerate}

Updating test code is often accompanied by the addition of new dependencies.
In the \textit{introduce required dependencies} section, the prompt instructs the LLM to add \texttt{import} statements only for newly introduced dependencies, avoiding duplicated imports. 
As for the response format, we instruct the LLM to respond only with Java code, beginning with new import statements and following with the updated test method, as illustrated in Listing ~\ref{lst:res_format}.

\textbf{\textit{Post-Processing.}}
After receiving outputs from the LLM, \appname performs a post-processing step. 
It parses the generated Java code to extract two key components: the updated test method and the \texttt{import} statements. 
These two components are then integrated into the test class during the subsequent iterative refinement phase, ensuring the completeness and executability of the generated test code.

\begin{lstlisting}[caption={An example of the response format}, label={lst:res_format}]
    // New import statements
    import com.example.NewDependency;
    import static org.mockito.Mockito.when;
    
    // Updated test methods
    @Test
    public void testFocalMethod() {
        // Updated test logic here
    }
\end{lstlisting}

\subsection{Iterative Refinement}
By emulating developers’ manual debugging processes, we design an error-type-aware iterative refinement module to further enhance the correctness of the updated test.
Guided by the findings from ChatTester~\cite{yuan2024evaluating} that symbol resolution and type error dominate compilation failures and most execution failures (85.5\%) are assertion errors, \appname classifies errors into three major categories, i.e., missing symbol, type mismatch, and assertion error. 
For each category, we design a targeted refinement strategy and extract the precise context required, rather than merely passing raw error messages back to the LLM.
The details are as follows.

\textbf{\textit{Error Analysis and Context Collection.}}
To ensure successful execution of the generated test case, \appname first replaces the $T_o$ with $T_u$ and inserts the new \texttt{import} statements generated by LLM into the test file. 
Then, \appname executes $T_u$ and performs error analysis based on the type of failure encountered:

\textbf{Compilation Failure:} 
    \textit{1) Cannot find symbol.} 
    This is the most common error encountered in the updated test code, which typically arises from missing or wrong \texttt{import} statements or references to undefined classes, methods, or variables within the test method. 
    For this failure case, \appname extracts the name and the corresponding line number of the missing symbol from the compiler's error message. 
    It then queries the language server to locate the symbol’s definition or the symbol that invokes or accesses it, such as \texttt{symbolName.symbolName}.
    Once the missing symbol’s definition is found, \appname constructs a prompt including the error location and definition to assist the LLM in refining the test case, as shown in Fig.~\ref{fig:motivation_iter}.

    \textit{2) Argument type mismatch (cannot be applied to given types).} 
    This error occurs when the argument types or number of arguments passed to a method do not match its definition. 
    In such cases, \appname queries the language server to retrieve the method’s definition, along with the error line, and uses these contexts to construct the prompt for repair.

    \textit{3) Other compilation errors.} 
    The above two error types cover the majority of compilation failures caused by LLM-generated test updates. 
    For other errors, either no additional context is needed (e.g., incompatible types), or the cases are too rare to justify custom handling.
    \appname simply extracts the error messages and locations and then passes them to the LLM.

\textbf{Test Failure:} 
Test failure means that $T_u$ fails to pass, which is typically caused by assertion errors.
Similar to compilation failure, \appname locates the test report and extracts context from it, including the expected and actual values of assertions, as well as the line number of the failed assertions. 
Then, \appname extracts the corresponding assertions from the test code. 
Together, the expected and actual values and the assertions are used to construct a prompt (which shares a similar design to the one shown in Fig.~\ref{fig:motivation_iter}) to guide the LLM in repairing $T_u$.

\textbf{\textit{The Validate-Repair Loop.}}
The entire process of validation and refinement is performed iteratively, with each round building upon the feedback from the previous failures to incrementally correct $T_u$.
This validate-repair loop continues until $T_u$ passes or the iteration limit is reached.
Our initial experiments show that most errors can be corrected within two rounds of repair. 
For the remaining cases, however, the LLM tends to fall into erroneous loops, making it unlikely to produce correct results even after additional attempts.
Thus, we limits the LLM to a maximum of two repair attempts. 

While this loop can successfully correct the majority of common issues, it may still fail in some scenarios. 
For example, when the focal method introduces new parameters and the associated testing logic is intricate, the LLM may struggle to construct correct test logic. 
To address such cases, \appname includes a fallback mechanism. 
If a correct pass is not achieved within the maximum number of repair iterations, the LLM is instructed to generate a \emph{basic answer}, which abandons the addition of new testing logic and instead applies a minimal modification strategy, such as assigning default values to the new parameters. 
This ensures that the test method can at least compile and run successfully.

\textbf{In summary}, by integrating LLMs with language servers and compiler feedback, \appname not only simulates the developer’s manual update process but also advances test maintenance from rule-based, static repair toward a dynamic, reasoning-driven update paradigm. 
The LLM-driven context collector enables semantic dependency resolution beyond simple syntactic diffs, while the error-type-aware refinement guides updates through targeted search rather than ad-hoc retries. Together, these components establish a generalizable framework that enhances both the correctness and adaptability of automated test updates.

\section{Dataset: \datasetname}
Existing datasets for test repair, such as \textsc{TaRBench}~\cite{target}, primarily focus on repairing failed test cases, while overlooking test cases that continue to pass after code changes but require additional test logic.
To address this limitation, we construct a new dataset, named \datasetname. 
\datasetname contains both broken test repair and unbroken test enhancement scenarios by collecting real-world production-test co-evolution samples.
Additionally, \datasetname incorporates the necessary runtime environment configurations to facilitate execution-based evaluation.
The dataset construction process involves two main stages: \emph{raw data collection} and \emph{data validation}.

\subsection{\textbf{Raw Data Collection}}
We follow two steps to collect the initial data:

\textbf{Project Collection.}
We first utilize the GitHub API to collect open-source Java projects. 
To facilitate compilation and the generation of test coverage reports, we only select projects that can be built with Maven and support JaCoCo. 
To ensure diversity and quality, we collect the following two categories of projects:
(1) 966 projects with 50 to 1000 stars on GitHub. 
These projects vary in structure and maturity, reflecting common development scenarios. 
(2) 64 projects from the top 500 most-starred Java projects on GitHub. 
These projects are typically well-maintained, actively developed, and follow standardized practices. 

\textbf{Sample Collection.}
First, we collect commits that involve the modifications of both production and test code.
Then, we employ the tree-sitter~\cite{tree-sitter} tool to parse the ASTs to find the modified production and test methods. 
If a modified test method invokes a modified production method, they are considered as a co-evolution sample.
According to Sun~\emph{et al.}~\cite{sun2023revisiting}, such modifications occurring within the same commit are most likely to represent the true production-test co-evolution.
To control the evaluation costs, we select 10 projects from each of the two categories that contain a sufficient number of qualifying samples.

In total, we collect 4,286 raw samples from 20 projects. 
Each sample includes both the original and updated pairs of production and test methods.

\subsection{\textbf{Data Validation}}
We conduct two steps to ensure the quality of the data:

\textbf{Test Execution}
To ensure the correctness of each test update, the original and updated tests must pass on the original and updated production code, respectively. 
Consequently, we execute the original and updated test methods before and after the code commit. 
We extract the required Java version from the \textit{pom.xml} file and determine the Maven version by consulting the \textit{README} or Maven wrapper configuration files.
If these versions are not specified, we automatically run the test methods across multiple versions.
A sample is considered valid only when both test methods pass successfully.
The runnable Java and Maven configurations are recorded in \datasetname.

\textbf{Manual Inspection}
We further manually check the samples to eliminate updates in which the production and test code are not semantically related. 
To control the manual inspection cost, we select 7 projects that have a moderate size and fast compilation speed from the 20 projects.
Additionally, since a single commit may contain multiple semantically redundant modifications~\cite{kim2009discovering}, we manually removed duplicate samples (i.e., same changes related to one changed focal method) to reduce data redundancy.

After the automated and manual filtering, we collected 195 method-level co-evolution samples of production and test code from seven selected projects.
Detailed information about the evaluation dataset \datasetname is shown in TABLE~\ref{tab:project_dataset}.
The dataset includes 165 broken test cases and 30 enhanced unbroken test cases, which reflect the realistic proportions in practical development scenarios.
And there are 92 method signature changes and 103 internal logic changes in \datasetname.

\begin{table}[tb]
  \footnotesize
  \centering
  \caption{Details of Projects in \datasetname}
  \label{tab:project_dataset}
  \begin{center}
    \begin{tabular}{lcc}
      \toprule
      \textbf{Name} & \textbf{Samples} & \textbf{Stars} \\
      \midrule
      \href{https://github.com/Aiven-Open/klaw}{Aiven-Open/klaw} & 50  & 161    \\
      \specialrule{0em}{1pt}{1pt}
      \href{https://github.com/alibaba/nacos}{alibaba/nacos} & 30  & 31k    \\
      \specialrule{0em}{1pt}{1pt}
      \href{https://github.com/apache/rocketmq}{apache/rocketmq} & 20  & 21.7k  \\
      \specialrule{0em}{1pt}{1pt}
      \href{https://github.com/apache/shenyu}{apache/shenyu} & 48  & 8.6k   \\
      \specialrule{0em}{1pt}{1pt}
      \href{https://github.com/OpenAPITools/openapi-generator}{OpenAPITools/openapi-generator} & 5   & 23.3k  \\
      \specialrule{0em}{1pt}{1pt}
      \href{https://github.com/prebid/prebid-server-java}{prebid/prebid-server-java} & 24  & 76     \\
      \specialrule{0em}{1pt}{1pt}
      \href{https://github.com/shred/acme4j}{shred/acme4j} & 18  & 541    \\
      \bottomrule
    \end{tabular}
  \end{center}
\end{table}

\section{Experimental Setup}
\label{expsetup}
In this section, we first introduce the baselines we used.
Then we describe the evaluation metrics in our experiments.
Finally, we present the experimental settings of our evaluation.

\subsection{Baselines}
We select the following methods as our baselines:

\textbf{\textsc{Ceprot}.}
\textsc{Ceprot}~\cite{ceprot} is a PLM-based methods that fine-tunes CodeT5~\cite{wang2021codet5} to automately update test cases.
We configure \textsc{Ceprot} using the settings that achieved the best results in its original experiments. 

\textbf{\textsc{TaRGET}.}
\textsc{TaRGET}~\cite{target} is also a PLM-based method that fine-tunes code language models to automate test repair.
We replicate \textsc{TaRGET} with the best settings in its experiments, which employs the CodeT5+~\cite{wang2023codet5opencodelarge} model.

\textbf{\textsc{Synter.}}
\textsc{Synter}~\cite{synter} is the first LLM-based test repair approach designed to repair broken test cases caused by syntactic breaking changes. 
Although it primarily focuses on test repair, its methodology is closely related to the test update task addressed in this study, as both aim to align test code with changes in production code. 

\textbf{\textsc{NaiveLLM.}}
\textsc{NaiveLLM} is a special baseline designed to evaluate the effectiveness of directly applying LLMs to the test update task. 
It simply provides the LLM with the code changes of the focal method along with the original test method.
By comparing \appname with NaiveLLM, we can better evaluate the effectiveness of our designs in context collection and iterative refinement.

\subsection{Evaluation Metrics}
Inspired by previous studies on automated test generation~\cite{yuan2024evaluating,wang2024hits}, test repair~\cite{synter}, and automated program repair~\cite{zhang2024reviewapm}, we employ the following three metrics to evaluate the effectiveness of \appname:
\emph{compilation pass rate}, \emph{test pass rate}, and \emph{test coverage}.

\textbf{Compilation Pass Rate (CPR)} refers to the proportion of generated test cases in the evaluation dataset that can compile successfully, reflecting the syntactic correctness of the generated test code. 

\textbf{Test Pass Rate (TPR)} refers to the proportion of updated tests that can successfully execute and pass. 
This metric is widely adopted in studies regarding test generation~\cite{yuan2024evaluating,chen2024chatunitest}.
The test pass rate assesses whether the generated tests conform to the current system behavior.

\textbf{Test Coverage (Cov.).}
Test coverage is a critical indicator of the effectiveness and quality of test suites. 
Common coverage metrics include line coverage and branch coverage, which evaluate the breadth and depth of behavioral validation performed by the test code, respectively. 

\subsection{Experimental Settings}
\textbf{LLM Selection.}
In the experiments, we evaluate three representative large language models (LLMs): Llama-3.3-70B-Instruct, GPT-4.1, and DeepSeek-V3. 
These models are selected to balance open-source and closed-source options, Llama-3.3-70B-Instruct and DeepSeek-V3 are open-source models, while GPT-4.1 is a closed-source commercial model. 
All three LLMs have a substantial number of parameters and large context windows, showing strong performance on diverse benchmarks, which is suitable for the test update task that requires powerful code understanding capabilities. 

\textbf{Implementation Details.}
We employ LangChain~\cite{langchain} to invoke LLM APIs and construct prompt templates. 
Llama-3.3-70B-Instruct is deployed locally using the vLLM~\cite{vllm} inference engine, whereas GPT-4.1 and DeepSeek-V3 are accessed via their respective official APIs. 
When evaluating the test cases, we executed them under the specific Java and Maven version recorded in \datasetname.
To mitigate the effect of stochasticity, we evaluate each LLM three times with a temperature of 0.1 on the dataset, and report the average compilation pass rate and test pass rate for \appname as well as the baselines.

\section{Evaluation Results}

We formulate the following research questions (RQs) to evaluate the performance of our proposed method \appname on the \datasetname dataset:

\begin{itemize}[leftmargin=*]
    \item \textbf{RQ1: Effectiveness of \appnamebold in Compilation and Test Pass Rate.} 
    How effective is \appname in test update in terms of \textit{compilation pass rate} (\textit{CPR}) and \textit{test pass rate} (\textit{TPR})?

    \item \textbf{RQ2: Effectiveness of \appnamebold in Test Coverage.}  
    How effective is \appname in test update in terms of \textit{test coverage (Cov.)}?

    \item \textbf{RQ3: Ablation Study.}  
    How effective are the core components of \appname in the test update?
\end{itemize}

\subsection{RQ1: Effectiveness of \appname in Compilation and Test Pass Rate.}
To evaluate the effectiveness of \appname in unit test update, we compare it with baseline methods under two metrics: \textbf{\textit{compilation pass rate (CPR)}} and \textbf{\textit{test pass rate (TPR)}}.
The results are shown in Table~\ref{tab:rq1-result}.

\begin{table}[t]
    \footnotesize
    \centering
    \caption{Compilation and Test Pass Rates across Models (195 samples)}
    \label{tab:rq1-result}
    \begin{tabular}{llcc}
        \toprule
        \textbf{Model} & \textbf{Method} & \textbf{CPR (\%)} & \textbf{TPR (\%)} \\
        \midrule
        & \textsc{Ceprot} & 20.5 & 9.7 \\
        & \textsc{TaRGET} & 45.1 & 41.0 \\
        \midrule
        \midrule
        \multirow{3}{*}{Llama-3.3-70B} 
            & \textsc{NaiveLLM} & 49.2 & 35.9 \\
            & \textsc{Synter}   & 65.1 & 55.3 \\
            & \appnamebold & \textbf{89.7} & \textbf{77.4} \\
        \midrule
        \multirow{3}{*}{GPT-4.1} 
            & \textsc{NaiveLLM} & 45.1 & 36.9 \\
            & \textsc{Synter}   & 79.5 & 68.2 \\
            & \appnamebold & \textbf{91.8} & \textbf{84.1} \\
        \midrule
        \multirow{3}{*}{DeepSeek-V3} 
            & \textsc{NaiveLLM} & 55.9 & 44.6 \\
            & \textsc{Synter}   & 78.5 & 66.7 \\
            & \appnamebold & \textbf{94.4} & \textbf{86.7} \\
        \bottomrule
    \end{tabular}
\end{table}

Regarding two PLM-based methods,
\textsc{Ceprot} achieves a low \textit{CPR} of 20.5\% and a \textit{TPR} of 9.7\%.
This is because CodeT5 has a limited maximum output length, resulting in most of its generated tests being incomplete.
Although \textsc{TaRGET} outperforms \textsc{Ceprot}, its performance is still significantly lower than \appname.
The reason is that \textsc{TaRGET} can only handle broken test cases with a single-hunk modification, also with a limited sequence length.
\textsc{TaRGET} can only generate valid outputs for 101 out of 195 cases.

As for LLM-based methods, \appname significantly outperforms \textsc{Synter} and \textsc{NaiveLLM} across all three LLMs on both metrics.
Compared to \textsc{Synter} (i.e., the best-performing baseline), \appname improves \textit{CPR} and \textit{TPR} by at least 12.3\% and 15.9\%, respectively.
\appname achieves the best performance when using DeepSeek-V3 as the underlying LLM, obtaining the highest \textit{CPR} of 94.4\% and \textit{TPR} of 86.7\%.
DeepSeek-V3 and GPT-4.1 achieve better performances than Llama-3.3-70B-Instruct, which indicates that our method's effectiveness is positively correlated with the performance of the underlying LLM.
\appname demonstrates the largest relative improvement over \textsc{Synter} on Llama-3.3-70B-Instruct, with enhancements of 24.6\% on \textit{\textit{CPR}} and 22.0\% on \textit{TPR}, respectively.
This suggests that our method yields more substantial improvements when applied to the LLM with relatively lower performance.

\find{\textbf{Answer to RQ1: }\appname achieves the best performance in both compilation and test pass rates, validating its superior performance in test update. 
We also find that the effectiveness of \appname is positively correlated with model capability and shows the most significant gains on weaker LLMs.
}

\subsection{RQ2: Effectiveness of \appname in Test Coverage.}
Since \appname achieves the highest compilation and test pass rates under the DeepSeek-V3 model in RQ1, we use the results from DeepSeek-V3 to evaluate the method's test coverage.
Specifically, we compare the \textit{\textbf{branch coverage (Branch Cov.)}} and \textbf{\textit{line coverage (Line Cov.)}} of three methods: \textsc{NaiveLLM}, \textsc{Synter}, \appname, along with the \textsc{Ground Truth} (i.e., the developer-update test cases from the \datasetname dataset). Table~\ref{tab:coverage-summary} presents the overall coverage, which is the average line or branch coverage across all test cases in the dataset, regardless of whether the tests pass or fail.

\begin{table}[t]
\caption{Overall Test Coverage under DeepSeek-V3}
\label{tab:coverage-summary}
\footnotesize
\centering
    \begin{tabular}{lcc}
    \toprule
    \textbf{Method} & \textbf{Branch Cov. (\%)} & \textbf{Line Cov. (\%)} \\
    \midrule
    \textsc{NaiveLLM} & 21.5 & 36.6 \\
    \textsc{Synter} & 33.1 & 53.9 \\
    \appname & 46.0 & 69.1 \\
    \textsc{Ground Truth} & 53.0 & 80.4 \\
    \bottomrule
    \end{tabular}
\end{table}

From the results, \appname achieves a branch coverage of 46.0\% and a line coverage of 69.1\%, which significantly outperforms \textsc{NaiveLLM} and \textsc{Synter} with improvements of at least 12\%.
The main reason why \appname's coverage is still lower than the \textit{Ground Truth} is that no matter how effective the method is, some test cases will inevitably fail and be recorded as 0\% coverage.

To enable a fairer comparison of coverage between the methods, we further evaluate the jointly-passed coverage by analyzing the test cases that both methods successfully executed.
Table~\ref{tab:coverage_synter} compares \appname with \textsc{Synter} on the jointly passed test cases. 
Table~\ref{tab:coverage_gt} compares \appname with the \textsc{Ground Truth} on the 169 test cases successfully passed by \appname.

\begin{table}[htbp]
\caption{Coverage Comparison Between \appname and \textsc{Synter}}
\label{tab:coverage_synter}
\footnotesize
\centering
    \resizebox{0.5\textwidth}{!}{
    \begin{tabular}{lcccc}
    \toprule
    \multirow{2}{*}{\textbf{Method}} & \multicolumn{2}{c}{\textbf{112 Jointly Passed Tests}} & \multicolumn{2}{c}{\textbf{Remaining 73 Tests}} \\
    \cmidrule(lr){2-3} \cmidrule(lr){4-5}
     & \textbf{Branch (\%)} & \textbf{Line (\%)} & \textbf{Branch (\%)} & \textbf{Line (\%)} \\
    \midrule
    \textsc{Synter} & 49.8 & 80.5 & 12.6 & 21.3 \\
    \appname & \textbf{54.8} & \textbf{81.1} & \textbf{38.3} & \textbf{58.0} \\
    \bottomrule
    \end{tabular}
    }
\end{table}

According to Table~\ref{tab:coverage_synter}, among the 112 test cases successfully passed by both methods, \appname achieves a branch coverage of 54.8\% and a line coverage of 81.1\%, both higher than \textsc{Synter}.
In the remaining 73 test cases, the gap becomes more pronounced: \appname achieves 38.3\% branch coverage and 58.0\% line coverage, significantly higher than \textsc{Synter}.
These results suggest that \appname performs better than \textsc{Synter} on easier cases (i.e., signature-related code changes), and significantly better on more complex or difficult ones that \textsc{Synter} fails to handle.

\begin{table}[t]
\caption{Coverage Comparison between \appname and Ground Truth}
\label{tab:coverage_gt}
\footnotesize
\centering
    \begin{tabular}{lcc}
    \toprule
    \textbf{Method} & \textbf{Branch Cov. (\%)} & \textbf{Line Cov. (\%)} \\
    \midrule
    \appname & \textbf{52.7} & \textbf{80.2} \\
    Ground Truth & 52.1 & 80.1 \\
    \bottomrule
    \end{tabular}
\end{table}

As shown in Table~\ref{tab:coverage_gt}, for the 169 test cases successfully passed by \appname, the average branch and line coverage reaches 52.7\% and 80.2\%, slightly higher than the \textsc{Ground Truth}. 
This indicates that \appname has achieved, and even slightly exceeded, the level of code coverage provided by manually written tests. 
A possible explanation is that the passed tests generated by \appname tend to systematically explore different branches and execution paths.

\find{\textbf{Answer to RQ2: }\appname outperforms \textsc{Synter} in both overall and jointly-pass test coverage, and surpasses \textsc{Ground Truth} in individual comparisons. 
It indicates that \appname is effective at verifying new functionality that \textsc{Synter} fails to cover, and sometimes even reaches or outperforms professional software developers.}

\subsection{RQ3: Ablation Study.}
\label{sec:ablation_study}
To investigate the individual contributions of the core components in \appname, we conducted an ablation study by evaluating different combinations. 
\appname consists of two key module: \textbf{\textit{Context Collection}} and \textbf{\textit{Iterative Refinement}}. 
Removing both modules results in the baseline \textsc{NaiveLLM} method. 
Accordingly, we compared the performance of \textsc{NaiveLLM}, \appname without context collection (\textsc{TestUpdater}\(_{wo}\)\textit{CC}), \appname without iterative refinement (\textsc{TestUpdater}\(_{wo}\)\textit{IR}), and the full \appname method in terms of compilation pass rate and test pass rate. 
The results are summarized in Table~\ref{tab:ablation_study}.

\begin{table}[htpb]
  \footnotesize
  \centering
  \caption{Ablation Study on the Effect of \appname Components}
  \label{tab:ablation_study}
    \begin{tabular}{ccc}
    \toprule
    \textbf{Method} & \textbf{CPR} (\%) & \textbf{TPR} (\%) \\
    \midrule
    \textsc{NaiveLLM} & 55.9 & 44.6 \\
    \textsc{TestUpdater}\(_{wo}\)\textit{IR} & 83.4 & 67.7 \\
    \textsc{TestUpdater}\(_{wo}\)\textit{CC} & 90.8 & 82.1 \\
    \appname          & 94.4 & 86.7 \\
    \bottomrule
  \end{tabular}
\end{table}

The ablation results indicate that \textsc{NaiveLLM} has relatively low performance, suggesting limited test update effectiveness.
Adding the context collection module (\textsc{TestUpdater}\(_{wo}\)\textit{IR}) yields an improvement of about 25\% in compilation and test pass rate.
Incorporating the iterative refinement module (\textsc{TestUpdater}\(_{wo}\)\textit{CC}) leads to a higher improvement, with the compilation and test pass rate rising around 35\%, highlighting the critical role of iterative refinement in improving the executability of generated code.

The comparatively smaller gain from the \textit{context collection} module can be attributed to the fact that the LLM introduces new, error-prone test logic to enhance coverage, which necessitates correction in the iterative refinement phase. 
By integrating these modules, \appname effectively enhances and updates obsolete test cases.

\find{\textbf{Answer to RQ3: } In summary, both \textit{context collection} and \textit{iterative refinement} modules contribute positively to \appname’s performance, with iterative refinement showing a more significant impact.}
\section{Discussion}

\subsection{Failure Analysis.}
To further investigate the limitations of \appname in test update tasks, we analyzed failed update cases and summarized the main causes of failure. 
The observed failure types can be categorized as follows:

\textbf{Function Parameter Mismatch.}  
When the focal method changes, especially involving additions or modifications of multiple parameters, the LLM may struggle to accurately infer the meaning or type of new parameters. 
This often leads to invocation errors in the generated test code, typically resulting in compilation failures (e.g., “cannot be applied to given types”) due to incorrect argument passing.

\textbf{Overly Complex Focal Method Changes.}  
If the modification to the focal method involves large-scale refactoring, complex control flows, or deep data dependencies, the LLM may fail to fully understand the change even when context is provided. 
As a result, the generated test logic may not adequately cover all new functional paths or assertion conditions, leading to test failures.

\textbf{Context Collection Failure.}  
During the context collection phase, \appname relies on the LLM to help identify key methods and classes relevant to the focal method changes. 
However, in some cases, the LLM returns elements not from the local source code but from external Maven dependencies. 
Since the language server can only index local project files, it fails to resolve these external references, resulting in missing context.
Meanwhile, the LLM may fail to correctly identify required methods or classes.

\subsection{Impact of Test Enhancement.}
To examine the trade-off between benefits and overhead of test enhancement, we modified our pipeline to let the LLM perform only test repair (repair-only) and compared it with the original \appname (repair-enhance). 
Table~\ref{tab:compair-enhance} summarizes the overall coverages and average time overheads (Avg. Time).
DeepSeek-V3 is used in this experiment because it achieved the best performance in RQ1.
Compared with the repair-only workflow, the full repair-enhance workflow achieves much higher coverage with little extra time.
This shows that enabling test enhancement is crucial for verifying new functionality and improving test quality with negligible efficiency loss.

\begin{table}[htpb]
\caption{Repair-Only vs. Repair-Enhance}
\label{tab:compair-enhance}
\footnotesize
\centering
    \begin{tabular}{lccc}
    \toprule
    \textbf{Workflow} & \textbf{Branch Cov. (\%)} & \textbf{Line Cov. (\%)} & \textbf{Avg. Time}\\
    \midrule
    repair-only & 35.3 & 54.7 & 70.7s\\
    repair-enhance & 46.0 & 69.1 & 71.8s \\
    \bottomrule
    \end{tabular}
\end{table}

\subsection{Evaluation on \textsc{TarBench} for Test Repair Comparison.}
\textsc{TarBench}~\cite{target} is a benchmark constructed by Yaraghi et al. to evaluate the test repair performance of \textsc{TaRGET}~\cite{target}, comprising 7,103 test instances focused on single-hunk test method repairs when the system under test evolves. 
In contrast, our \datasetname benchmark focuses on general test method updates in response to changes in the focal methods. 
To facilitate a comprehensive comparison with existing test repair methods, we also evaluated \appname on \textsc{TarBench} and compared its performance with \textsc{TaRGET}.

Considering the large size of \textsc{TarBench}, we utilized a locally deployed model Llama-3.3-70B-Instruct as the underlying LLM of \appname to accommodate the potential high costs. 
For \textsc{TaRGET}, we used the best results reported in its original paper. 
\textsc{TaRGET} generates 40 outputs per test case, and we ran \appname five times per test case. 
A case is considered passed if any of these outputs succeed.
As presented in Table~\ref{tab:tarbench}, \appname achieves performance comparable to \textsc{TaRGET} on \textsc{TarBench}, demonstrating its effectiveness in dedicated single-hunk test repair scenarios. 
We anticipate that by leveraging more advanced LLMs, such as GPT-4.1 or DeepSeek-V3, \appname’s performance can be further enhanced (as presented in RQ1).
Besides, in contrast to \textsc{TaRGET} that requires task-specific pretraining, \appname showcases a more general approach, which requires only the off-the-shelf LLMs but achieves comparable performance on single-hunk test repair scenarios and much better performance on general test update scenarios. 

\begin{table}[htpb]
\caption{Test Repair Performance on \textsc{TarBench}}
\label{tab:tarbench}
\footnotesize
\centering
    \begin{tabular}{lcc}
    \toprule
    \textbf{Method} & \textbf{CPR (\%)} & \textbf{TPR (\%)} \\
    \midrule
    \textsc{TaRGET} & 89.6 & 80.0 \\
    \appname & 87.4 & 83.6 \\
    \bottomrule
    \end{tabular}
\end{table}

\subsection{Test Update vs. Test Generation}
To update test cases, one can also leverage test generation tools to generate brand-new unit tests instead of updating the existing ones.
To investigate the effectiveness of this method in just-in-time test update, we evaluated the state-of-the-art test generation approach HITS~\cite{wang2024hits} on our dataset.
All settings (e.g., prompts and post-processing strategies) were kept the same as HITS. 
The only difference is the underlying LLM, which we used DeepSeek-V3 with temperature set to 0.1, consistent with our experimental setup (see Section~\ref{expsetup}.C) to ensure fairness.
Among the 195 test cases, HITS achieved a compilation pass rate of 33.3\% and a test pass rate of 18.9\%, substantially lower than existing LLM-based test update baselines and our approach (see Table~\ref{tab:rq1-result}).

\begin{figure*}[t]
  \centering
  \includegraphics[width=0.88\textwidth]{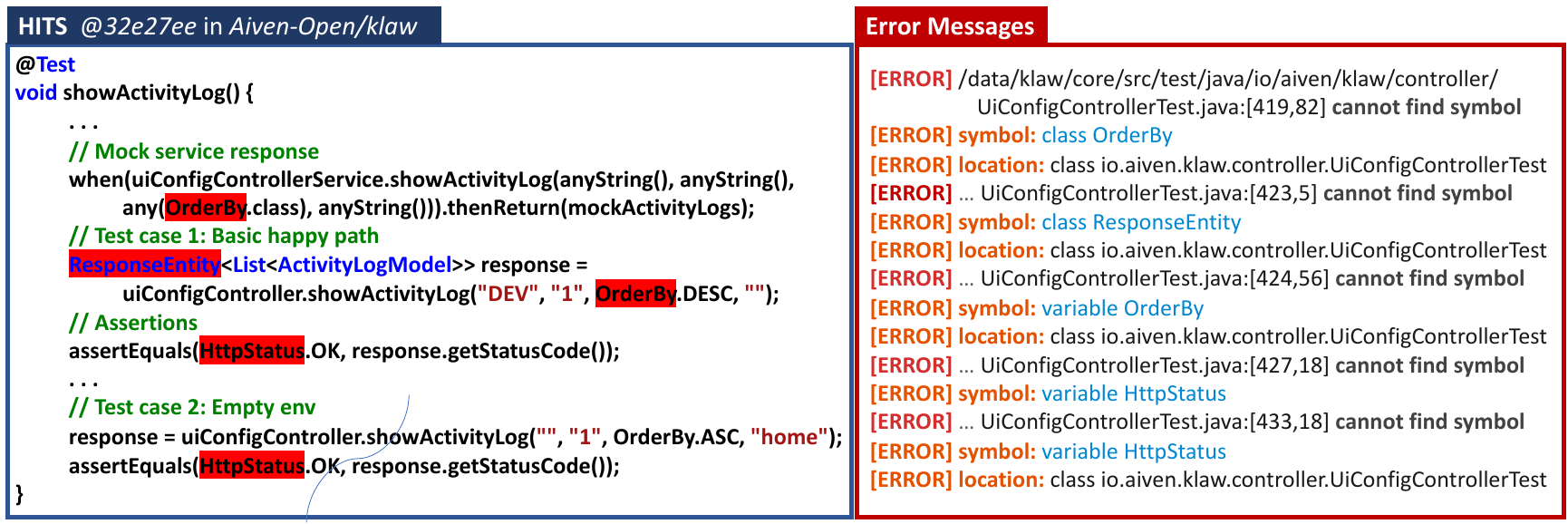}
  \caption{An Import Error Example of HITS.}
  \label{fig:hits_error}
  \vspace{-10pt}
\end{figure*}

We observed that most failures in HITS stem from import errors.
Fig.~\ref{fig:hits_error} presents the test generated by HITS for the same case shown in Fig.~\ref{fig:motivation_iter} (which is correctly updated by our approach). 
Symbols highlighted in red were not imported, resulting in “cannot find symbol” errors.
Although HITS iteratively attempts to fix errors based on the reported messages, these messages (see the right part of Fig.~\ref{fig:hits_error}) only indicate the line numbers of the missing symbols, without providing their package paths for importing.
Consequently, the errors remain unrecoverable regardless of the number of iterations.

\subsection{Efficiency}
The time overhead of \appname consists of three main components: (1) initializing the language server, which typically takes from a few seconds to a few minutes; (2) invoking the LLM; and (3) compiling and executing the updated test methods. 
Based on our measurements, updating a single test case takes an average of 71.8 seconds and 4 LLM invocations.
Although our method is less time-efficient than \textsc{Synter} (10–30 seconds) due to the necessity of compiling and executing tests in each refinement iteration, it prioritizes correctness and offers potential for future improvements.

In terms of token usage, each update consumes an average of 2,197 tokens, approximately 1,000 more than \textsc{Synter} due to more frequent invocations and larger context. 
However, this additional cost leads to noticeably better performance.

\subsection{Threats to Validity}

\textbf{\textit{Flaky Test Detection.}}
To eliminate the potential impact of flaky tests on our evaluation and avoid introducing instability to developers, we executed each test case 10 times under identical conditions. 
A test is considered flaky if it yields inconsistent outcomes across runs. 
Our results show that all \textsc{Ground Truth} and \appname-generated test cases were consistent, indicating no flakiness.

\textbf{\textit{Internal Validity.}}
The dataset used in the experiments is sourced from publicly available code repositories.
While it aims to cover common types of changes in real-world software evolution, it may not fully represent all test update scenarios. 
Additionally, the \datasetname dataset may carry a risk of data leakage.
However, considering that the same LLMs are used in the baseline models, the relative performance improvement of \appname in test generation remains valid from a controlled variable perspective.

\textbf{\textit{External Validity.}}
The generalization of our method may be limited since \appname currently targets the Maven-JUnit framework, which may not represent projects with different build systems or programming languages. 
However, the Maven-JUnit framework is commonly used in real-world software.
Future work could extend \appname to support more diverse ecosystems.

\section{Related Works}

\subsection{Automated Test Repair}

Automated test repair aims to fix broken test cases caused by changes in the corresponding production code. 
Early approaches relied on heuristics and program analysis.
ReAssert~\cite{daniel2011reassert} applied strategies such as literal replacement and assertion adjustment. 
Symbolic Test Repair~\cite{daniel2010test} leveraged symbolic execution to modify failing literals. 
TestCareAssistant~\cite{TestCareAssistant} combined static and dynamic analysis to adjust inputs and assertions, while Test Fix~\cite{xu2014using} formulated test repair as a search problem solved via genetic algorithms.

More recent work explores learning-based and LLM-based methods. 
\textsc{Ceprot}~\cite{ceprot} employs a transformer-based model to identify and update obsolete test cases in a two-stage pipeline, showing strong performance across Java projects. 
\textsc{TaRGET}~\cite{target} formulates test repair as a code-to-code translation task, using rule-based algorithms to collect context and finetuning models to achieve 66.1\% exact match accuracy on the \textsc{TarBench} dataset.
\textsc{Synter}~\cite{synter} is the first LLM-based method, focusing on syntactic-breaking changes.
\textsc{Synter} leverages static analysis and a neural reranker to enhance the performance of LLMs and outperform \textsc{Ceprot} and \textsc{NaiveLLM} in text similarity and accuracy.
Unlike these methods, our approach not only repairs broken tests but also enhances non-broken tests to improve coverage.

\subsection{Automated Test Generation}

Automated test generation aims to reduce manual effort in writing test cases. Traditional tools such as EvoSuite~\cite{evosuite}, Randoop~\cite{randoop}, and Symstra~\cite{symstra} rely on evolutionary algorithms, random testing, or symbolic execution.
A series of learning-based methods has recently been proposed.
AthenaTest~\cite{athenatest} leverages BART~\cite{lewis2019bart} to generate test cases, while Atlas~\cite{watson2020learning} and Tufano \emph{et al.}~\cite{tufano2022generating} focus on generating assertion statements. 
However, these methods often lack guarantees of syntactic correctness or executability~\cite{yang2022surveydl}.

Recent research has explored the integration of LLMs into automated test generation. 
ChatTester~\cite{yuan2024evaluating} is the first LLM-based approach, demonstrating superior performance over EvoSuite and AthenaTest.
Chen \emph{et al.}~\cite{chen2024chatunitest} proposed a practical framework, ChatUniTest, employing a generate-verify-repair loop to improve accuracy and coverage. 
SymPrompt~\cite{symprompt} introduces path constraint prompts inspired by symbolic execution to guide LLMs for higher code coverage.
HITS~\cite{wang2024hits} applies code slicing to divide complex methods into smaller units, improving both line and branch coverage.
Compared with our work, we focus on repairing and enhancing existing tests in the process of software evolution, whereas these methods are directed at generating new tests for a specific method.

\section{Conclusion and future works}

In this paper, we propose \appname, an LLM-based approach for automated just-in-time unit test updates that can effectively update outdated unit tests caused by production code changes. 
\appname leverages LLM for collecting adequate and precise context, guides test updates with carefully crafted prompts, and introduces an iterative refinement mechanism to increase the correctness of the LLM-generated tests.
Notably, \appname not only repairs broken tests but also enhances test quality by validating newly introduced logic.
To support evaluation, we construct \datasetname, a benchmark dataset containing real-world co-evolution scenarios of production and test methods. 
Experimental results show that \appname outperforms all the baselines \textsc{Synter}, \textsc{NaiveLLM}, \textsc{Ceprot}, and \textsc{TaRGET} across compilation pass rate and test pass rate.
Meanwhile, \appname also outperforms LLM-based baselines \textsc{Synter} and \textsc{NaiveLLM} in test coverage.
Future work includes expanding support for additional languages and frameworks, optimizing efficiency for industrial-scale deployment.
Overall, this paper provides a novel approach and practical reference for continuous test maintenance during software evolution and demonstrates the potential of LLMs in automated test update.

\section*{Acknowledgment}
This research/project was supported by the National Natural Science Foundation of China (No.62202420), Zhejiang Provincial Natural Science Foundation of China (No.LZ25F020003), and the Fundamental Research Funds for the Central Universities (No.226-2025-00067).

\balance

\end{document}